\documentclass[conference]{IEEEtran}

\usepackage{amsmath}
\usepackage{amssymb}
\usepackage{graphics}
\usepackage{graphicx}
\usepackage{subcaption}
\usepackage{breqn}
\usepackage[lined,boxed,commentsnumbered,linesnumbered, ruled]{algorithm2e}
\usepackage{fixltx2e}		

\usepackage[font=footnotesize]{caption}

\newcommand{\trans}{^\mathsf{T}}

\newcommand{\cA}{{\mathsf{A}}}
\newcommand{\cB}{{\mathsf{B}}}
\newcommand{\nA}{n_{\mathsf{A}}}
\newcommand{\nB}{n_{\mathsf{B}}}
\newcommand{\CA}{C_{\mathsf{A}}}
\newcommand{\CB}{C_{\mathsf{B}}}

\newcommand{\D}{\boldsymbol{D}}
\newcommand{\A}{\boldsymbol{A}}
\newcommand{\C}{\boldsymbol{C}}
\renewcommand{\c}[2]{\ensuremath{\mathcal {#1}_{#2}}}		

\newtheorem{theorem}{Theorem}

\newtheorem{definition}{Definition}
\newtheorem{proposition}{Proposition}
\newtheorem{example}{Example}

\newcommand\scalemath[2]{\scalebox{#1}{\mbox{\ensuremath{\displaystyle #2}}}}

\IEEEoverridecommandlockouts

\begin{document}

\title{A Family of Erasure Correcting Codes with Low Repair Bandwidth and Low Repair Complexity}


\author{
\IEEEauthorblockN{Siddhartha Kumar$^\dag$, Alexandre Graell i Amat$^\dag$, Iryna Andriyanova$^\ddag$, and Fredrik Br\"annstr\"om$^\dag$}
\IEEEauthorblockA{$\dag$Department of Signals and Systems, Chalmers University of Technology, Gothenburg, Sweden\\
  $\ddag$ETIS Laboratory, ENSEA/University of Cergy-Pontoise/CNRS, Cergy-Pontoise, France}
              \thanks{This work was partially funded by the Swedish Research Council under grant \#2011-5961. Siddhartha Kumar is now with the University of Bergen and the Simula Research Lab, Norway. }\vspace*{-1cm}
}


\maketitle

\begin{abstract}

We present the construction of a new family of erasure correcting codes for distributed storage that yield low repair bandwidth and low repair complexity. The construction is based on two classes of parity symbols. The primary goal of the first class of symbols is to provide good fault tolerance, while the second class facilitates node repair, reducing the repair bandwidth and the repair complexity. We  compare the proposed codes with other codes proposed in the literature.
\end{abstract}


\section{Introduction}

Distributed storage (DS) uses a network of interconnected inexpensive storage devices (referred to as storage nodes or simply nodes) to store data reliably over long periods of time. Reliability against node failures (commonly referred to as \textit{fault tolerance}) is achieved by means of erasure correcting coding.  Furthermore, when a node fails,  a new node needs to be added to the DS network and populated with data to maintain the initial state of reliability. The problem of \textit{repairing} a failed node is known as the \textit{repair problem}.


Classical maximum distance separable (MDS) codes are optimal in terms of the fault tolerance/storage overhead tradeoff. However, the repair of a failed node requires the retrieval of large amounts of data from a large subset of nodes. Therefore, in the recent years, the design of erasure correcting codes that reduce the cost of repair has attracted a great deal of attention. Pyramid codes \cite{hua07} were one of the first code constructions that addressed this problem. In particular, they aim at reducing the number of nodes that need to be contacted to repair a failed node, known as the repair access. Other codes that reduce the repair access are the local reconstruction codes (LRCs) \cite{hua12}, and the locally repairable codes \cite{sat13,pap12}. 

Other code constructions aim at reducing the repair bandwidth, defined as the amount of information that needs to be read from the DS network to repair a failed node. Among them, we can mention minimum disk I/O repairable (MDR) codes \cite{wan14}, Zigzag codes \cite{tam13} and piggyback codes \cite{ras13}. Piggybacking consists of adding carefully chosen linear combinations of data symbols (called piggybacks) to the parity symbols of a given erasure correcting code. This results in a lower repair bandwidth at the expense of a lower erasure correcting capability with respect to the original code.

In this paper, we propose a family of erasure correcting codes that achieve low repair bandwidth and low repair complexity. In particular, we propose a systematic code construction based on two classes of parity symbols. Correspondingly, there are two classes of parity nodes. The first class of parity nodes, whose primary goal is to provide erasure correcting capability, is constructed using an MDS code modified by applying specially designed piggybacks to some of its code symbols. The second class of parity nodes is constructed using a block code whose parity symbols are obtained with simple additions. This class of parity nodes does not have the purpose to bring any additional erasure correcting capability, but to facilitate node repair at low repair bandwidth and low repair complexity. In the paper, we compare the proposed codes with MDR codes, Zigzag codes, piggyback codes and LRCs \cite{hua12}, in terms of repair bandwidth and repair complexity.


Notation:  We define the operator $(a+b)_k\triangleq(a+b)\mod k$. The Galois field of order $q^p$ is denoted by $\mathbb F_{q^p}$.



	
\section{Code Construction}
\label{sec:CodeConstruction}

\begin{figure}[!t]
\centering
\includegraphics{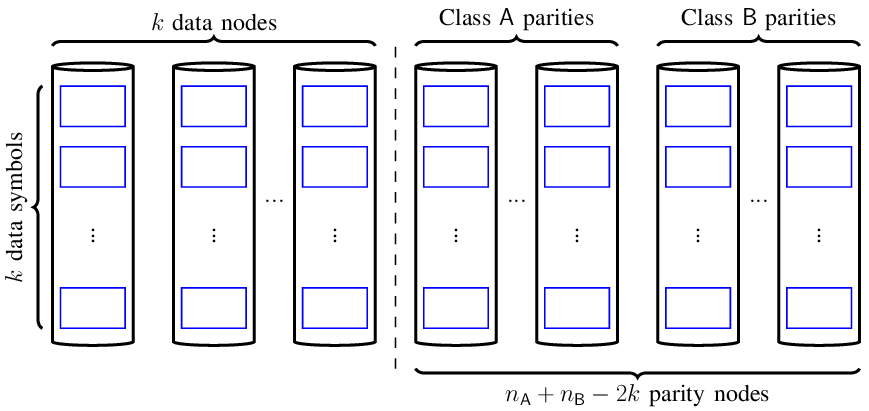}
\vspace{-3ex}
\caption{System model.}
\label{Fig:SystemModel}
\vspace{-3ex}
\end{figure}

We consider the distributed storage system depicted in Fig.~\ref{Fig:SystemModel}. There are $k$ data nodes, each containing a very large number of data symbols over $\mathbb F_{q^p}$. As we shall see in the sequel, the proposed code construction works with blocks of $k$ data symbols per node. Thus, without loss of generality, we assume that each node contains $k$ data symbols. We denote by $d_{i,j}$, ${i,j=0,\ldots,k-1}$, the $i$th data symbol in the $j$th data node. We say that the data symbols form a $k\times k$ \textit{data array} $\D$, where $d_{i,j}=[\D]_{i,j}$. For later use, we also define the set of data symbols $\mathcal D\triangleq\{d_{i,j}\}$. Further, there are $n-k$ parity nodes each storing $k$ parity symbols. We denote by $p_{i,j}$, $i=0,\ldots,k-1$, $j=k,\ldots,n-1$, the $i$th parity symbol in the $j$th parity node, and define the set $\mathcal P_j$ as the set of parity symbols in the $j$th parity node. The set of all parity symbols is denoted by $\mathcal P\triangleq \cup_j \{\mathcal{P}_j\}$. We say that the data and parity symbols form a $k\times n$ \textit{code array} $\C$, where $c_{i,j}=[\C]_{i,j}$. Note that $c_{i,j}=d_{i,j}$ for $i,j=0,\ldots,k-1$ and $c_{i,j}=p_{i,j}$ for $i=0,\ldots,k-1$, $j=k,\ldots,n-1$.

Our main goal is to construct codes that yield low repair bandwidth and low repair complexity of a single failed systematic node. To this purpose, we construct a family of systematic $(n,k)$ codes consisting of two different classes of parity symbols. Correspondingly, there are two classes of parity nodes, referred to as Class $\cA$ and Class $\cB$ parity nodes, as shown in Fig.~\ref{Fig:SystemModel}. Class $\cA$ and Class $\cB$ parity nodes are built using an $(\nA,k)$ code and an $(\nB, k)$ code, respectively, such that $n=\nA+\nB-k$. In other words, the parity nodes of the $(n,k)$ code\footnote{With some abuse of language we refer to the nodes storing the parity symbols of a code as the parity nodes of the code.} correspond to the parity nodes of Class $\cA$ and Class $\cB$ codes. The primary goal of Class $\cA$ parity nodes is to achieve good erasure correcting capability, while the purpose of Class $\cB$ nodes is to yield low repair bandwidth and low repair complexity. In particular, we focus on the repair of data nodes. The repair bandwidth (in bits) per node, denoted by $\gamma$, is proportional to the average number of symbols (data and parity) that need to be read to repair a data symbol, denoted by $\lambda$. More precisely, let $\beta$ be the number of symbols per node\footnote{For our code construction, $\beta=k$, but this is not the case in general.}. Then,
\begin{align}
\lambda = \frac{\gamma}{\nu \beta},
\label{eq:lambda}
\end{align}
where $\nu=\lceil\log_2 q^p\rceil$ is the size (in bits) of a symbol. $\lambda$ can be interpreted as the repair bandwidth normalized by the size (in bits) of a node. Therefore, in the rest of the paper we will use $\lambda$ to refer to the normalized repair bandwidth.

The main principle behind our code construction is the following. The repair is performed one symbol at a time. After the repair of a data symbol is accomplished, the symbols read to repair that symbol are cached in the memory. Therefore, they can be used to repair the remaining data symbols at no additional read cost. The proposed codes are constructed in such a way that the repair of a new data symbol requires a low additional read cost (defined as the number of additional symbols that need to be read to repair the data symbol), so that $\lambda$ (hence $\gamma$) is reduced. Since we will often use the concepts of read cost and additional read cost in the remainder of the paper, we define them in the following.
\begin{definition}
The \textit{read cost} of a symbol is the number of symbols that need to be read to repair the symbol. The \textit{additional read cost} of a symbol is the additional number of symbols that need to be read to repair the symbol, considering that other symbols are already cached in the memory (i.e., have been read to recover some other data symbols previously).
\end{definition}

\section{Class $\cA$ Parity Nodes}
\label{Sec:ClassA}

Class $\cA$ parity nodes are constructed using a modified $(\nA, k)$ MDS code, $k+ 2 \le \nA < 2k$, over $\mathbb F_{q^p}$. In particular, we start from an $(\nA, k)$ MDS code and apply piggybacks \cite{ras13} to some of the parity symbols. The construction of Class $\cA$ parity nodes is performed in two steps as follows.
\begin{enumerate}
\item[1)] Encode each row of the data array using an $(\nA,k)$ MDS code (the same for each row). The parity symbol $p^{\cA}_{i,j}$ is\footnote{We use the superscript $\cA$ to indicate that the parity symbol is stored in a Class $\cA$ parity node.}
\begin{align}
p^{\cA}_{i,j}=\sum_{l=0}^{k-1} \alpha_{l,j}d_{i,l},~j=k,\ldots {\color{blue},} \nA-1,
\label{eq:pij}
\end{align}
where $\alpha_{l,j}$ denotes a coefficient in $\mathbb F_{q^p}$.
Store the parity symbols in the corresponding row of the code array. Overall, $k(\nA-k)$ parity symbols are generated.
\item[2)] Modify some of the parity symbols by adding piggybacks. Let $\tau$, $1\leq\tau\leq \nA-k-1$, be the number of piggybacks introduced per row. The parity symbol $p_{i,u}$ is obtained as
\begin{align}
\label{eq:piu}
p^{\cA}_{i,u}= p^{\cA}_{i,u}  + d_{(i+u-\nA+\tau+1)_k, i},
\end{align}
where $ u=\nA-\tau,\ldots,\nA-1$ and the second term in the summation is the piggyback.  
\end{enumerate}

The \emph{fault tolerance} (i.e., the number of node failures that can be tolerated) of Class $\cA$ codes is given in the following theorem.
\begin{theorem}
An $(\nA,k)$ Class $\cA$ code with $\tau$ piggybacks per row can correct a minimum of $\nA-k-\tau+1$ node failures.
\label{th:ECC}
\end{theorem}
	\begin{IEEEproof}
		The proof is given in the appendix.
	\end{IEEEproof}

\begin{figure}[t]
		\centering
\includegraphics{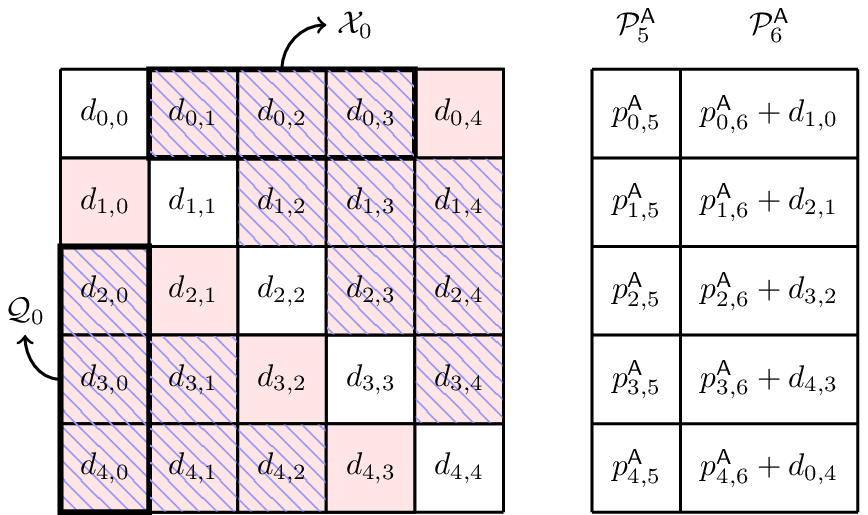}
		\vspace{-2ex}
		\caption{A $(7, 5)$ Class $\cA$ code with $\tau=1$ constructed from a $(7, 5)$ MDS code. 
		$\mathcal{P}_{5}^{\cA}$ and $\mathcal{P}_{6}^{\cA}$ are the parity nodes. For each row $j$, colored symbols belong to $\mathcal{R}_{j}$.}
		\label{fig2}
		\vspace{-3ex}
\end{figure}

When a failure of a data node occurs, Class $\cA$ parity nodes are used to repair $\tau+1$ of the $k$ failed symbols. The Class $\cA$ parity symbols are constructed in such a way that, when node $j$ is erased, $\tau+1$ data symbols in this node can be repaired reading the (non-failed) $k-1$ data symbols in the $j$th row of the data array and $\tau+1$
parity symbols in the $j$th row of Class $\cA$ nodes (see also Section~\ref{sec:Decoding}). For later use, we define the set $\mathcal R_j$ as follows.
\begin{definition}
For $j=0,\ldots,k-1$, define the set $\mathcal{R}_{j}$ as $\mathcal{R}_{j}=\{d_{j,(j+1)_k},d_{j,(j+2)_k}, \cdots,d_{j,(j+k-1)_k} \}$.
\end{definition}

The set $\mathcal{R}_{j}$ is the set of $k-1$ data symbols that are read from row $j$ to recover $\tau+1$ data symbols of node $j$ using Class $\cA$ parity nodes.
\begin{example}
An example of Class $\cA$ code is shown in Fig.~\ref{fig2}. One can verify that the code can correct any 2 node failures. For each row $j$, the set \c{R}{j} is indicated in red color.  For instance, $\mathcal{R}_{0}=\{d_{0,1},d_{0,2},d_{0,3},d_{0,4}\}$.
\end{example}

The main purpose of Class $\cA$ parity nodes is to provide good erasure correcting capability. However, the use of piggybacks helps also in reducing the number of symbols that need to be read to repair the $\tau+1$ symbols of a failed node that are repaired using Class $\cA$ code, as compared to MDS codes. The remaining $k-\tau-1$ data symbols of the failed node can also be recovered from Class $\cA$ parity nodes, but at a high symbol read cost. Hence, the idea is to add another class of parity nodes, namely Class $\cB$ parity nodes, in such a way that these symbols can be recovered with lower read cost.

\section{Class $\cB$ Parity Nodes}
\label{Sec:ClassB}

Class $\cB$ parity nodes are obtained using an $(\nB,k)$ linear block code over $\mathbb F_{q^p}$ to encode the $k\times k$ data symbols of the data array, i.e., we use the $(\nB,k)$ code $k$ times. This generates $(\nB-k)\times k$ Class $\cB$ parity symbols, $p^{\cB}_{i,u}$, $i=0,\ldots,k-1$, $u=\nA,\ldots,n-1$.

\begin{definition}
For $j=0,\ldots,k-1$, define the 
	set $\mathcal{Q}_{j}$ as 
	\begin{align}
		\mathcal{Q}_{j}&= \{d_{(j+\tau+1)_k, j},d_{(j+\tau+2)_k, j}, \dotsm, d_{(j+k-1)_k,j}\}.
	\end{align}
\end{definition}

Assume that data node $j$ fails. It is easy to see that the set $\mathcal Q_j$ is the set of $k-\tau-1$ data symbols that are not recovered using Class $\cA$ parity nodes. 
\begin{example}
\label{ex:Q0R0}
For the example in Fig.~\ref{fig2}, the set $\mathcal{Q}_{j}$ is indicated by hatched symbols for each column $j$, $j=0,\ldots, k-1$. For instance, $\mathcal{Q}_{0}=\{d_{2,0},d_{3,0},d_{4,0}\}$. 
\end{example}

For later use, we also define the following set.
\begin{definition}
For $j=0,\ldots,k-1$, define the set $\mathcal{X}_{j}$ as
\begin{align}
\mathcal{X}_{j}&=\{d_{j,(j+1)_k},d_{j,(j+2)_k}, \dotsm, d_{j,(j+k-\tau-1)_k}\}.
\end{align}
\end{definition}
Note that $\mathcal{X}_{j}=\mathcal{R}_{j}\cap\{\cup_{l} \mathcal{Q}_{l}\}$.

\begin{example}
\label{ex:X0}
For the example in Fig.~\ref{fig2}, the set $\mathcal{X}_i$ is indicated by hatched symbols for each row $i$. For instance, $\mathcal{X}_{0}=\mathcal{R}_{0}\cap \{ \mathcal{Q}_{0}\cup \mathcal{Q}_{1}\cup \mathcal{Q}_{2}\cup \mathcal{Q}_{3}\cup \mathcal{Q}_{4}\}=\{d_{0,1},d_{0,2},d_{0,3}\}$.	
\end{example}

The purpose of Class $\cB$ parity nodes is to allow recovering of the data symbols in $\mathcal{Q}_{j}$, $j=0,\ldots,k-1$, at a low additional read cost. Note that after recovering $\tau+1$ symbols using Class $\cA$ parity nodes, the data symbols in $\mathcal{R}_{j}$  are already stored in the decoder memory, therefore they are accessible for the recovery of the remaining $k-\tau-1$ data symbols using Class $\cB$ parity nodes without the need of reading them again. The main idea is based on the following proposition.
\begin{proposition}
\label{prep:MainIdea}
If a Class $\cB$ parity symbol $p^{\cB}$ is the sum of one data symbol $d\in \mathcal{Q}_{j}$ and a number of data symbols in $\mathcal{X}_{j}$, then the recovery of $d$ comes at the cost of one additional read (one should read parity symbol $p^{\cB}$).
\end{proposition}

This observation is used in the construction of Class $\cB$ parity nodes (see Section~\ref{sec:big-example} below) to reduce the normalized repair bandwidth, $\lambda$. In particular, we add $k-\tau-1$ Class $\cB$ parity nodes which allow to reduce the additional read cost of all $k(k-\tau-1)$ data symbols in all $\mathcal{Q}_{j}$'s to $1$. (The addition of a single Class $\cB$ parity node allows to recover one new data symbol in each $\mathcal{Q}_{j}$, $j=0,\ldots,k-1$, at the cost of one additional read).



%
	\begin{figure*}
		\begin{subfigure}[H]{0.4\columnwidth}
			\centering
			\includegraphics{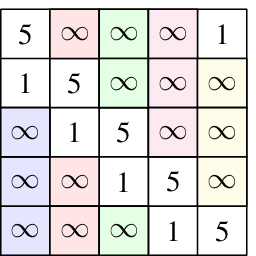}
			\caption{Initial.}
			\label{fig3a}
		\end{subfigure}
		\begin{subfigure}[H]{0.4\columnwidth}
			\centering
			\includegraphics{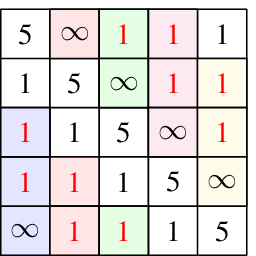}
			\caption{Step 1c.}
			\label{fig3b}
		\end{subfigure}
		\begin{subfigure}[H]{0.4\columnwidth}
			\centering
			\includegraphics{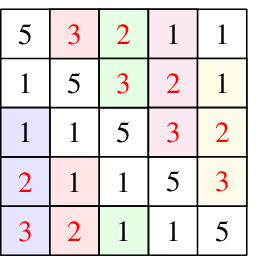}
			\caption{Step 1d.}
			\label{fig3c}
		\end{subfigure}
		\begin{subfigure}[H]{0.4\columnwidth}
			\centering
			\includegraphics{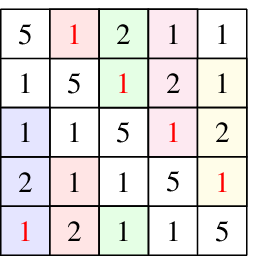}
			\caption{Step 2b.}
			\label{fig3d}
		\end{subfigure}
		\begin{subfigure}[H]{0.2\textwidth}
			\centering
			\includegraphics{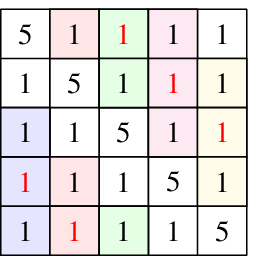}
			\caption{Step 3b.}
			\label{fig:fig3e.}	
		\end{subfigure}

		\caption{Update of $\A$ during the construction of Class $\cB$ parity nodes for the example in Section~\ref{sec:big-example}.
		The updates of $a_{i,j}$ after each step are highlighted in red color. The shaded symbols in column $j$ denote the set $\mathcal{Q}_{j}$, while the shaded symbols in row $i$ denote the set $\mathcal{X}_{i}$.
}
\label{fig3}
\vspace{-3ex}
\end{figure*}

In order to describe the code construction, we define the function $\mathrm{read}(d, p^\cB)$ 
as follows. 
\begin{definition}
\label{def:read}
Consider a Class $\cB$ parity node and let ${\mathcal P}^{\cB}$ denote the set of parity symbols in this node. Also, let $d\in \mathcal{Q}_{j}$ for some $j$ and $p^{\cB}\in{\mathcal P}^{\cB}$ be $p^{\cB}=d+\sum_{d'\in \mathcal D'} d'$, where $\mathcal D'\subset\mathcal D$, i.e., the parity symbol  $p^{\cB}$ is the sum of $d$ and a subset of other data symbols.
Then,
	\begin{align}
		\mathrm{read}(d,p^{\cB})=|\breve{\mathcal D}\backslash \mathcal{X}_{j}|,
	\end{align}
where $\breve{\mathcal D}=\{\mathcal D'\cup d\}$.
\end{definition} 

For a given data symbol $d$, the function $\mathrm{read}(d,p^{\cB})$ gives the additional number of symbols that need to be read to recover $d$ (considering the fact that some symbols are already cached in the memory).


\subsection{Construction Example}

In the following, we propose a recursive algorithm for the construction of Class $\cB$ parity nodes. To ease understanding, we introduce the algorithm through an example.

\label{sec:big-example}
We construct a $(10,5)$ code starting from the $(7,5)$ Class $\cA$ code in Fig.~\ref{fig2}. In particular, we construct $k-\tau-1=3$ Class $\cB$ parity nodes, so that the additional number of reads to repair each of the remaining failed $k-\tau-1=3$ symbols (after recovering $\tau+1=2$ symbols using Class $\cA$ parity nodes) is $1$. With some abuse of notation, we denote these parity nodes by $\mathcal{P}_{7}^{\cB}$, $\mathcal{P}_{8}^{\cB}$, and $\mathcal{P}_{9}^{\cB}$.

Denote by $\A$, $a_{i,j}=[\A]_{i,j}$, a temporary matrix of read values for the respective data symbols $d_{i,j}$. After Class $\cA$ decoding,
\begin{align}
	a_{i,j}=
	\begin{cases}
		\infty & \text{if } d_{i,j}\in \{\cup_t \mathcal{Q}_t\}\\
		k & \text{if } i=j\\
		1 &  \text{otherwise},
	\end{cases}	
	\label{Eq:A}
\end{align}
where $t=0,\ldots,k-1$.
For our example, $\A$ after Class $\cA$ decoding is given in Fig.~\ref{fig3}(a).
Our algorithm operates on the $a_{i,j}$s whose initial value is $\infty$ and aims to obtain the lowest possible values for these $a_{i,j}$s under the given number of Class $\cB$ parity nodes. This is done in a recursive manner as follows. 

\begin{itemize}
\item[\textbf{1.}] \textbf{Construct the first parity node, $\mathcal{P}_{7}^{\cB}$.}
\end{itemize}
\begin{enumerate}
	\item[1a] For each symbol $d_{i,j}$ define the set $\mathcal{\tilde{D}}_{i,j}\triangleq\{d_{(i+s)_k,(j+s)_k}\}_{s=0}^{k-1}$.
	\item[1b] Start with the elements in $\mathcal{Q}_{0}$. Pick an element $d_{i,0}\in \mathcal{Q}_{0}$ such that $a_{i,0}=\infty$, and $d_{0,i}\in \mathcal{X}_{0}\backslash \mathcal{\tilde{D}}_{i,0}$. For instance, we take $d_{2,0}$.
	\item[1c] For $t=0,\ldots,k-1$ compute
	\begin{align}
	\label{eq:pt7}
		p^{\cB}_{t,7}=d_{(i+t)_k,t}+d_{t,(i+t)_k}
	\end{align}
	and update the respective $a_{i,0}$ and $a_{0,i}$,
	\begin{align}
		a_{(i+t)_k,t}=a_{t,(i+t)_k}=\mathrm{read}(d_{(i+t)_k,t},p^{\cB}_{t,7}).
	\end{align}
	The resulting matrix $\A$ is shown in Fig.~\ref{fig3}(b). There are still entries $a_{i,j}=\infty$ that need to be handled.  
	\item[1d] For $t=0,\ldots,k-1$ update
	\begin{align}
		p^{\cB}_{t,7}=p^{\cB}_{t,7}+d_{t,(i'+t)_k},
	\end{align}
	where $d_{0,i'}\in \mathcal{X}_{0}$ and $a_{0,i'}=\infty$ after step 1b. Update $\A$ accordingly (see Fig.~\ref{fig3}(c)). Note that the read values $a_{(i+t)_k,(j+t)_k}$ have not worsened. This comes from the fact that the new added data symbol belongs to the corresponding set $\mathcal X$ and is already cached in the memory. Thus, the additional read cost is $0$. On the other hand, the values $a_{(j+t)_k,(i+t)_k}$ increase.
\end{enumerate}
\begin{itemize}
	\item[\textbf{2.}] \textbf{Construct the second parity node, $\mathcal{P}_{8}^{\cB}$.}
\end{itemize}
\begin{enumerate}
	\item[2a] Pick an element $d_{i,0}\in \mathcal{Q}_{0}$ such that the corresponding $a_{i,j}$ is maximal. In our example, this is $d_{4,0}$ because $a_{4,0}=3$.
	\item[2b] For $t=0,\ldots,k-1$, do the following. Pick an element $d_{t,(u+t)_k}\in \mathcal{X}_{t}\backslash\mathcal{\tilde{D}}_{i,j}$ such that for all $d_{i',j'}\in \breve{\mathcal{D}}$, $\textrm{read}(d_{i',j'},p_{t,8})\le a_{i',j'}$, where $p^{\cB}$ is set to $p^{\cB}_{t,8}=d_{(i+t)_k,t}+d_{t,(u+t)_k}$. For our example, we choose $d_{0,2}$. Note that the only other option, $d_{0,3}$, is not a good choice as the new additional read cost would increase from 1 to 2. If such $d_{t,(u+t)_k}$ does not exist, set $p^{\cB}_{t,8}=d_{(i+t)_k,t}$.

Update $\A$. The updated matrix is shown in Fig.~\ref{fig3}(d).
\end{enumerate}
\begin{itemize}
	\item[\textbf{3.}] \textbf{Construct $\mathcal{P}_{9}^{\cB}$.} 
\end{itemize}
\begin{enumerate}
	\item[3a]  Pick an element $d_{i,0}\in \mathcal{Q}_{0}$ such that the corresponding $a_{i,0}$ is maximal. In our example, this is $d_{3,0}$.
	\item[3b]  For $t=0,\ldots,k-1$, do the following. $p^{\cB}_{t,9}=d_{(i+t)_k,t}$.
			Update $\A$. The resulting $\A$ has value $k$ for all diagonal elements and $1$ elsewhere (Fig.~\ref{fig3}(e)).
\end{enumerate}

\begin{figure}[!t]
	\centering
	\includegraphics{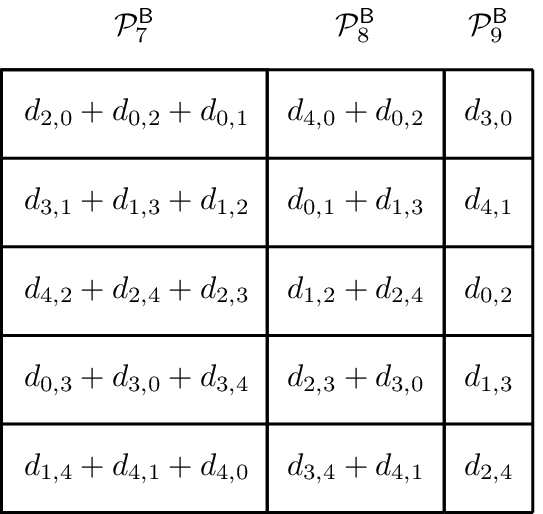}
	\caption{Class $\cB$ parity nodes for the data nodes in Fig.~\ref{fig2}.}
	\vspace{-3ex}
	\label{Fig:ClassB}
\end{figure}
The Class $\cB$ parity nodes $\mathcal{P}_{7}^{\cB}$, $\mathcal{P}_{8}^{\cB}$, and $\mathcal{P}_{9}^{\cB}$ are shown in Fig.~\ref{Fig:ClassB}.

A general version of the algorithm to construct Class $\cB$ parity nodes is given in Appendix~\ref{app:Algorithm}.

\subsection{Discussion of the Construction Example}

The construction of Class $\cB$ parity nodes starts by selecting an element $d_{i,j}$ of a given $\mathcal{Q}_{j}$ such that $a_{i,j}=\infty$ and  $d_{j,i}\in \mathcal{X}_{j} \backslash \mathcal{\tilde{D}}_{i,j}$ (for simplicity, as in the example, we can start with $j=0$). The first parity symbol of $\c{P}{7}$ after step 1c is therefore $p_{0,7}=d_{i,0}+d_{0,i}$, and the remaining parity symbols are obtained as in (\ref{eq:pt7}). By Proposition~\ref{prep:MainIdea} the additional read cost of $d_{i,j}$ (after step 1c) is $1$. The reason for selecting $d_{j,i}\in \mathcal{X}_{j} \backslash \mathcal{\tilde{D}}_{i,j}$ is due to the fact that, again by Proposition~\ref{prep:MainIdea}, its additional read cost is also $1$. We remark that for each $d_{i,j}\in \mathcal{Q}_{j}$ it is not always possible to select $d_{j,i}\in \mathcal{X}_{j} \backslash \mathcal{\tilde{D}}_{i,j}$ and set $p_{j,7}=d_{i,j}+d_{j,i}$. This is the case when $k<2(\tau+1)$. If $d_{j,i}\in \mathcal{X}_{j} \backslash \mathcal{\tilde{D}}_{i,j}$ does not exist, then we select $d_{j,t}\in \mathcal{X}_{j}\backslash\mathcal{\tilde{D}}_{i,j}$ (see Appendix~\ref{app:Algorithm}). In this case, the additional read cost of $d_{j,t}$ (after step 1c) is $>1$.

In general, step 1d has to be performed $|\mathcal{Q}_{j}| -2$  times, corresponding to the number of entries $a_{i,j}=\infty$ per column of $\A$.

Adding $k-\tau-1$ Class $\cB$ nodes allows to reduce the additional read cost for all data symbols in all $\mathcal{Q}_{j}$ to $1$ (see Fig.~\ref{fig3}(e)). However, this comes at the expense of a reduction in the code rate, i.e.,
the storage overhead is increased. In the example, $k-\tau-1=3$ Class $\cB$ parity nodes need to be introduced, which reduces the code rate from $R=5/7$ to $R=5/10=1/2$. If a lower storage overhead is required, Class $\cB$ parity nodes can be \textit{punctured}, starting from the last parity node (for the example, nodes $\mathcal{P}_{9}^{\cB}$, $\mathcal{P}_{8}^{\cB}$, and $\mathcal{P}_{7}^{\cB}$ are punctured in this order), at the expense of an increased repair bandwidth. If all Class $\cB$ parity nodes are punctured, we would remain only with Class $\cA$ parity nodes and the repair bandwidth corresponds to that of the Class $\cA$ code.  Thus, our code construction gives a family of rate-compatible codes which trades off between repair bandwidth and storage overhead: adding more Class $\cB$ nodes reduces the repair bandwidth but increases the storage overhead.

\subsection{Repair of a Single Node Failure: Decoding Schedule}
\label{sec:Decoding}
The repair of a failed systematic node, proceeds as follows. First, $\tau+1$ symbols are repaired using Class $\cA$ parity nodes. Then, the remaining symbols are repaired using Class $\cB$ parity nodes. With a slight abuse of language, we will refer to the repair of symbols using Class $\cA$ and Class $\cB$ parity nodes as the decoding of Class $\cA$ and Class $\cB$ codes, respectively. Suppose that node $j$ fails. Decoding is as follows.
\begin{itemize}
\item \textbf{Decoding of Class $\cA$ code}. To reconstruct the failed data symbol in the $j$th row of the code array, $k$ symbols ($k-1$ data symbols and $p^{\cA}_{j,k}$) in the $j$th row are read. These symbols are now cached in the memory. We then read  the $\tau$ piggybacked symbols in the $j$th row. By construction (see (\ref{eq:piu})), this allows to repair $\tau$ failed symbols, at the cost of an additional read each.
\item \textbf{Decoding of Class $\cB$ code}. Each remaining failed data symbol $d_{i,j}\in\mathcal Q_j$ is obtained by reading a Class $\cB$ parity symbol whose corresponding set $\breve{\mathcal D}$ (see Definition~\ref{def:read}) contains $d_{i,j}$. In particular, if several Class $\cB$ parity symbols $p^{\cB}_{i',j'}$ contain $d_{i,j}$, we read the parity symbol with largest index $j'$. This yields the lowest additional read cost.
\end{itemize}


\section{Code Characteristics and Comparison}
\label{Sec:Comparison}

\begin{table*}[h]
	\centering
	\caption{\footnotesize Comparison of $(n,k)$ codes that aim at reducing repair bandwidth. The repair bandwidth and the repair complexity are normalized per symbol, while the encoding complexity is given per row in the code array.  Note that for MDR codes $n=k+2$.}
	\label{Tab:Summary}
	\vspace{-1ex}
	\begin{tabular}{@{}p{1.9 cm}cp{2.0 cm}ccc@{}}
		\hline	
			& $\beta$ & \textbf{Fault Tolerance} & \textbf{Norm. Repair Band.} & \textbf{Norm. Repair Compl.} & \textbf{Enc. Complexity}\\
			\hline\\[-5pt]
			MDS & $1$ & \centering $n-k$ & $k$ & $O((k-1)\nu+k\nu^2)$ & $O((n-k)((k-1)\nu)+k\nu^2)$\\[5 pt]
			LRC \cite{hua12} & $1$ & \centering $r+1$ & $\frac{k}{n-k-r}$ & $O((\lceil\frac{k}{n-k-r}\rceil-1)\nu)$ & \scalebox{0.72}{$rO((k-1)\nu+k\nu^2)+(n-k-r)O((\lceil\frac{k}{n-k-r}\rceil-1)\nu)$}\\[5 pt]
			MDR \cite{wan14} & $2^k$ & \centering $2$ & $\frac{k+1}{2}$ & $O((k-1)\nu)$ & $O((k-1)\nu)$\\[5 pt]
			Zigzag \cite{tam13} & $(n-k)^{k-1}$ & \centering $n-k$ & $\frac{n-1}{n-k}$ &$O((k-1)\nu+k\nu^2)$ & $O((n-k)((k-1)\nu)+k\nu^2)$\\[5 pt]
			Piggyback \cite{ras13} & $2$ & \centering $1$ & \scalebox{0.9}{$\frac{(k-t_r)(k+t)+t_r(k+t_r+\ell-2)}{2k}$} & -- & --\\[5 pt]
			Proposed Codes & $k$ & \centering $\ge n-\nB-\tau+1$ & \scalebox{0.9}{$<\frac{k+\tau+(k-\tau-1)^2}{k}$} & $C_{\textrm{R}}/k$ & $C_{\textrm{E}}$\\[5 pt]
 		\hline
	\end{tabular}  
	\vspace{-2ex}
\end{table*}

In this section we characterize some different properties of the codes presented in Sections~\ref{Sec:ClassA} and \ref{Sec:ClassB}. 

\subsection{Fault Tolerance}
\label{sec:tolerance}
The fault tolerance of the Class $\cA$ code depends on the MDS code used in its construction and $\tau$, as stated in Theorem \ref{th:ECC}. Hence, our proposed code has also fault tolerance $f\ge\nA-k-\tau+1$. Since $1 \le \tau \le \nA -k-1$, our codes have a fault tolerance of at least $2$. 

\subsection{Normalized Repair Bandwidth}

According to Section \ref{sec:Decoding}, to repair the first $\tau+1$ symbols in a failed node requires that $k-1$ data symbols plus $\tau+1$ Class $\cA$ parity symbols are read. The remaining $k-\tau-1$ data symbols in the failed node are repaired by reading the Class $\cB$ parity symbols. As seen in Section \ref{Sec:ClassB}, the parity symbols in the first Class $\cB$ parity node are constructed from sets of data symbols of cardinality $|\mathcal{Q}_{j}|=k-\tau-1$. Therefore, to repair each of the $k-\tau-1$ data symbols in this set requires to read at most $k-\tau-1$ symbols. The remaining Class $\cB$ parity nodes are constructed from fewer symbols than $k-\tau-1$. An upper bound on the normalized repair bandwidth is therefore $\lambda<(k+\tau+(k-\tau-1)^2)/k$. It is observed that when $\tau$ increases, the fault tolerance reduces while $\lambda$ improves.

\subsection{Repair Complexity of a Failed Node}

We first consider the complexity of elementary arithmetic operations of elements of size $\nu=\lceil\log_2 q^p\rceil$ in $\mathbb F_{q^p}$. An  addition requires $O(\nu)$ and multiplication requires $O(\nu^2)$. The term inside $O(\cdot)$ denotes the number of elementary binary additions. To repair the first symbol requires $k$ multiplications and $k-1$ additions. To repair the following $\tau$ symbols require an additional $\tau k$ multiplications and additions. The final $k-\tau-1$ symbols require at most $k-\tau-2$ additions, since Class $\cB$ parity symbols are constructed as the sum of at most $k-\tau-1$ data symbols. The repair complexity of one failed node is therefore
\begin{align}
	\scalemath{0.95}{
	C_{\textrm{R}}=O((k-1)\nu+k\nu^2)+O(\tau k(\nu+\nu^2))+O((k-\tau-2)^2\nu)}.
\end{align}
The first two terms correspond to the Class $\cA$ code while the last term corresponds to the Class $\cB$ code.
	
\subsection{Encoding Complexity}
\label{sec:EncComp}

The encoding complexity of the $(n,k)$ code, $C_{\textrm{E}}$, is the sum of the encoding complexities of the two codes. The generation of each of the $\nA-k$ Class $\cA$ parity symbols in one row of the code array, $p^{\cA}_{i,j}$ in \eqref{eq:pij}, requires $k$ multiplications and $k-1$ additions. Adding data symbols to $\tau$ of these parity symbols according to \eqref{eq:piu} requires an additional $\tau$ additions. The encoding complexity of the Class $\cA$ code is therefore 
\begin{align}
	\CA= O((\nA-k)(k\nu^2+(k-1)\nu))+ O(\tau\nu).
\end{align}

According to Section \ref{Sec:ClassB}, the parity symbols in the first Class $\cB$ parity node are constructed as the sum of $k-\tau-1$ data symbols, and each parity symbol in the subsequent parity nodes uses one less data symbol. Therefore, the encoding complexity of the Class $\cB$ code is
	\begin{align}
		\CB=\sum_{i=1}^{n-\nA} O((k-\tau-1-i)\nu).
\end{align}
Finally, $C_{\textrm{E}}=\CA+\CB$.


\subsection{Code Comparison}

Table~\ref{Tab:Summary} provides a summary of the characteristics of different codes proposed in the literature as well as the codes constructed in this paper.\footnote{The variables $t, t_r$ and $r$ in Table~\ref{Tab:Summary} are defined in \cite{ras13} and \cite{hua12} respectively. The definition of $\ell$ comes directly from $r$ that is defined in \cite{ras13}.} In the table, column 2 reports the value of $\beta$ (see \eqref{eq:lambda}) for each code construction. For our code, $\beta=k$, unlike for MDR and Zigzag codes, for which $\beta$ grows exponentially with $k$. This implies that our codes require less memory to cache data symbols during repair. The fault tolerance $f$, the normalized repair bandwidth $\lambda$, the normalized repair complexity, and the encoding complexity, discussed in the previous subsections, are reported in columns 3, 4, 5, and 6, respectively.
	
\begin{figure}[t]
	\centering
	\includegraphics{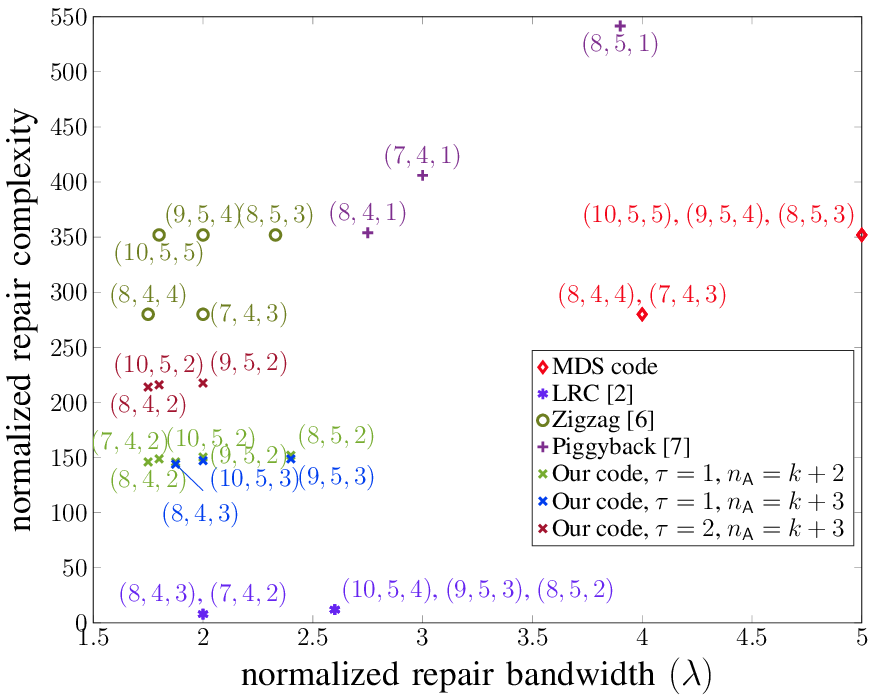}
	\caption{Comparisons of different codes $(n,k,f)$ with $\nu=8$.} 
	\label{fig5}
	\vspace{-3ex}
\end{figure}

In Fig.~\ref{fig5}, we compare our codes with other codes in the literature. In particular, the figure plots the normalized repair complexity of $(n,k,f)$ codes over $\mathbb F_{2^8}$ ($\nu=8$) versus their normalized repair bandwidth $\lambda$. In contrast to the bounds for the repair bandwidth and complexity reported in Table \ref{Tab:Summary}, Fig.~\ref{fig5} contains the exact number of integer additions. 

The best codes for a DS system should be the ones that achieve the lowest repair bandwidth and have the lowest repair complexity. As seen in Fig.~\ref{fig5}, MDS codes have both high repair complexity and repair bandwidth, but they are optimal in terms of fault tolerance for a given $n$ and $k$. Zigzag codes achieve the same fault tolerance and high repair complexity as MDS codes, but at the lowest repair bandwidth. At the other end, LRCs yield the lowest repair complexity but a higher repair bandwidth and worse fault tolerance than Zigzag codes. Piggyback codes have a repair bandwidth between that of Zigzag and MDS codes, but with a higher repair complexity and worse fault tolerance. For a given storage overhead, our proposed codes have better repair bandwidth than MDS codes, Piggyback codes and LRCs, and equal or similar repair bandwidth than Zigzag codes. Furthermore, they yield lower repair complexity as compared to MDS, Piggyback and Zigzag codes. However, the benefits in terms of repair bandwidth and/or repair complexity with respect to MDS and Zigzag codes come at a price of a lower fault tolerance.

\section{Conclusion}

In this paper, we constructed a new class of codes that achieve low repair bandwidth and low repair complexity for a single node failure. The codes are constructed from two smaller codes, Class $\cA$ and $\cB$, where the former focuses on the fault tolerance of the code, and the latter focuses on reducing the repair bandwidth and complexity. Our proposed codes achieve better repair complexity than Zigzag codes and Piggyback codes and better repair bandwidth than LRCs, but at the cost of slightly lower fault tolerance. A side effect of such a construction is that the number of symbols per node that needs to be encoded grows linearly with the code dimension. This implies that our codes are suitable for memory constrained DS systems as compared to Zigzag and MDR codes, for which the number of symbols per node increases exponentially with the code dimension. 


\appendices
\section{Proof of Theorem \ref{th:ECC}}

Each row in the code array contains $\nA-k-\tau$ parity symbols based on the MDS construction (i.e., parity symbols without piggybacks). Using these symbols, one can recover $\nA-k-\tau$ data symbols in that row and, thus, $\nA-k-\tau$ failures of systematic nodes. In order to prove the theorem, we need to show that by using piggybacked parity symbols $p_{i,u}$, $i=0, \ldots, k-1$, in some parity node, $u$, it is possible to correct one arbitrary systematic node failure. To do this, let us consider the system of linear equations $\boldsymbol{Gd}\trans=\boldsymbol{p}\trans$, representing the set of parity equations to compute $p_{i,u}$s where $u=\nA-\tau$. 
In other words, $\boldsymbol{d}=(d_{0,0}, \ldots, d_{0,k-1}, d_{1,0}, \ldots, d_{k-1,k-1}) $, $\boldsymbol{p} = (p_{0,u}, \ldots, p_{k-1,u})$, and 
$\boldsymbol G$ is given by 
\begin{align}
	\label{Eq:G}
	\boldsymbol G &=\scalemath{0.8}{\left(\begin{array}{cccccc}
		\boldsymbol a & \boldsymbol u_0 & \boldsymbol 0 & \boldsymbol 0&\dotso & \boldsymbol 0\\
		\boldsymbol 0 & \boldsymbol a & \boldsymbol u_1 & \boldsymbol 0&\dotso & \boldsymbol 0\\
		\boldsymbol 0 & \boldsymbol 0 & \boldsymbol a & \boldsymbol u_2&\dotso & \boldsymbol 0\\
		\vdots & \vdots & \vdots & \vdots & \ddots & \vdots \\
		\boldsymbol u_{k-1} & \boldsymbol 0 & \boldsymbol 0 & \boldsymbol 0&\dotso & \boldsymbol a\\
	\end{array}\right)}
\end{align}
where $\boldsymbol a=(\alpha_{0,u},\ldots, \alpha_{k-1,u})$, $\boldsymbol u_i$ is a vector of length $k$ with one at position $i$ and zeros elsewhere, and $\boldsymbol 0$ is the all-zero vector of size $k$.
Now, assume a systematic node $r$ has failed. In order to repair it, 
we need to solve the following subsystem of linear equations $\boldsymbol{G}'\boldsymbol w\trans=\boldsymbol p\trans$, in which $\boldsymbol w=(d_{0,r}, \ldots, d_{k-1,r})$ and $\boldsymbol G'$ is a $k \times k$ submatrix of $\boldsymbol G$ such that: 
a) its diagonal elements are all $\alpha_{r,u}$; 
b) it has 1 at row $r$ and column $(r+1)_k$; c) all other entries are 0. 
Note that $\boldsymbol{G}'$ is full rank. Therefore, one arbitrary data symbol can be corrected and, hence, the erasure correcting capability of Class $\cA$ code is at least $\nA-k-\tau+1$, which completes the proof.

\section{Algorithm to Construct Class $\cB$ Parity Nodes}
\label{app:Algorithm}

\begin{algorithm}[t]
	\DontPrintSemicolon
	\SetArgSty{textup}
	\SetKwRepeat{Do}{do}{while}
	\SetKwInput{Init}{Initialization}
	\Init{\\ \Indp $\forall i,j=0,\ldots,k-1$
	       \\ \Indp $a_{i,j}$ as defined in (\ref{Eq:A}) 
		  \\ $\mathcal{\tilde{D}}_{i,j}\triangleq\{d_{(i+s)_k,(j+s)_k}\}_{s=0}^{k-1}$ 
		  \\ \Indm $max\_itr=k-\tau-2$}
	\For(\\ \tcp*[h]{construct $k-\tau-1$ nodes}){$\omega\leftarrow \nA$ \KwTo $\nA+k-\tau-2$}{
	choose $d_{i,0}\in \mathcal{Q}_{0}$ s.t. $a_{i,0}$ is max $\&\&$ $d_{0,i}\in \mathcal{X}_{0}\backslash\mathcal{\tilde{D}}_{i,0}$ \\
	\lIf{$d_{0,i}\not\in \mathcal{X}_{0}\backslash\mathcal{\tilde{D}}_{i,0}$}{choose $d_{i,0}\in \mathcal{Q}_{0}$ s.t. $a_{i,0}$ is max}
	$p_{0,\omega}^\cB=d_{i,0}$ 
	\\
		\For{$t\leftarrow1$ \KwTo $k-1$}{
			$p_{t,\omega}^\cB=d_{(i+t)_k,t}$	
		}
		\For{$itr\leftarrow 1$ \KwTo $max\_itr$}{
			$temp=p_{0,\omega}^\cB+d_{0,i}$\\
			\eIf{$itr=1$ $\&\&$ $d_{0,i}\in \mathcal{X}_{0}\backslash\mathcal{\tilde{D}}_{i,0}$ $\&\&$  $\text{read}(d_{0,i},temp)< a_{0,i}$ }{
				$i'\leftarrow i$\\
				$p_{0,\omega}^\cB=temp$\\
				$a_{0,i'}=a_{i',0}=\text{read}(d_{0,i'},p_{0,\omega}^\cB)=1$\\
				\For{$t\leftarrow1$ \KwTo $k-1$}{
					$p_{t,\omega}^\cB=p_{t,\omega}^\cB+d_{t,(i'+t)_k}$\\
					$a_{t,(i'+t)_k}=\text{read}(d_{t,(i'+t)_k},p_{t,\omega}^\cB)$\\
					$a_{(i+t)_k,t}=\text{read}(d_{(i+t)_k,t},p_{t,\omega}^\cB)$		
				}
					
			}{
				\If{$\exists d_{0,i'}\in \mathcal{X}_{0}\backslash\mathcal{\tilde{D}}_{i,0}$ $\&\&$ $\text{read}(d_{0,i'},p_{0,w}^\cB)\leq a_{0,i'}$ $\&\&$ $a_{0,i'}>1$ }{
					$p_{0,\omega}^\cB=p_{0,\omega}^\cB+d_{0,i'}$\\
					$a_{0,i'}=max\_itr+1$\\
					$a_{0,i}=\text{read}(d_{0,i},p_{0,\omega}^\cB)$\\
					\For{$t\leftarrow1$ \KwTo $k-1$}{
						$p_{t,\omega}^\cB=p_{t,\omega}^\cB+d_{t,(i'+t)_k}$\\
						$a_{t,(i'+t)_k}=max\_itr+1$\\
						$a_{t,(i+t)_k}=\text{read}(d_{t,(i+t)_k},p_{t,\omega}^\cB)$
					}
				}
			}
		}
		$max\_itr\leftarrow max\_itr-1$
	}
	\caption{Construction of Class $\cB$ parity nodes}
	\label{alg:Algorithm1}
\end{algorithm}

We give an algorithm to construct $k-\tau-1$ Class $\cB$ parity nodes in the order $\mathcal{P}_{\nA}^\cB, \mathcal{P}_{\nA+1}^\cB,\ldots,\mathcal{P}_{\nA+k-\tau-2}^\cB$. This results in the construction of $(k-\tau-1)k$ parity symbols $p_{t,j}^{\cB}$. The algorithm is given in Algorithm~\ref{alg:Algorithm1}. Consider the construction of the parity symbols of parity node $\mathcal{P}_{\nA}$. The algorithm constructs first the parity symbol $p^\cB_{0,\nA}$ as the sum of an element $d_{i,0}\in\mathcal{Q}_{0}$ and $max\_itr$ elements in $\mathcal{X}_{0}$. Then, the other parity symbols $p^\cB_{t,\nA}$, $t>0$, are constructed as the sum of an element $d_{(i+t)_k,t}\in\mathcal{Q}_{t}$ and $max\_itr$ elements in $\mathcal{X}_{t}$, i.e., following a specific pattern. The remaining parity nodes are constructed in a similar way, with the only difference that the number of elements added from the sets $\mathcal{X}_{t}$, $max\_itr$, varies for each parity node.
The construction of the parity symbols $p^\cB_{t,j}$ depends on the choice of the symbols in the sets $\mathcal{Q}_{t}$ and $\mathcal{X}_{t}$. Assume that a parity symbol $p^\cB_{0,j}$ is constructed. The data symbols involved in $p^\cB_{0,j}$ are picked as follows.

\begin{itemize}
	\item Choice of a data symbol in $\mathcal{Q}_{0}$: Select a symbol $d_{i,0}\in \mathcal{Q}_{0}$ such that the corresponding $a_{i,0}$ is maximum and there exists $d_{0,i}\in \mathcal{X}_{0}\backslash\tilde{\mathcal D}_{i,0}$ (lines 2 and 3 in the algorithm). If the latter does not exists, then select $d_{i,0}$ such that $a_{i,0}$ is maximum. Such a $d_{i,0}$ always exist.
\item Choice of $max\_itr$ data symbols in $\mathcal{X}_{0}$: Select $max\_itr$  symbols $d_{0,i'}\in \mathcal{X}_{0}\backslash\tilde{\mathcal D}_{i,0}$ such that $a_{0,i'}>1$ and its additional read cost does not increase (line 20 in the algorithm).
If such a condition is not met, then the symbol $d_{0,i'}$ is not used in the construction of the parity symbol.
\end{itemize}

After the construction of each parity symbol, the corresponding entry of matrix $\boldsymbol{A}$ is updated.

\bibliographystyle{IEEEtran}

\end{document}